\documentclass[aps,prb,twocolumn,superscriptaddress,showpacs,floatfix]{revtex4}

\usepackage{graphicx,amsmath,amssymb}

\begin{document}

\title{Local distribution approach to disordered binary alloys}

\author{A.~Alvermann}
\author{H.~Fehske}
\affiliation{Institut f\"ur Physik, Ernst-Moritz-Arndt Universit\"at
  Greifswald, 17489 Greifswald, Germany}

\begin{abstract}
We study the electronic structure of the binary alloy and
(quantum) percolation model. Our study is based on a self-consistent
scheme for the distribution of local Green functions.
We obtain detailed results for the density of states,
from which the phase diagram of the binary alloy model is constructed,
and discuss the existence of a quantum percolation threshold.
\end{abstract}


\maketitle

\section{Introduction}\label{intro}

Many solids,
like alloys or doped semiconductors,
form crystals consisting of two or more chemical species.
In contrast to amorphous solids they possess, at least
approximately, a regular lattice whose sites are randomly occupied by
the different components.
Understanding their electronic structure is 
an important task which we will address here through an
approach that allows for a comprehensive description of such
substitutionally disordered systems. 
This approach, which we call the local distribution (LD) approach, 
considers the local density of states (LDOS) $\rho_i(\omega)$,
which is a quantity of primary importance in systems
with prominent local interactions or scattering, 
e.g. in the formation of local magnetic moments.
What makes the LD approach `non-standard' is that it directly deals with the
\emph{distribution} of the LDOS in the spirit that Anderson
introduced in his pioneering work.~\cite{anderson58}
While the LDOS is directly related to the amplitude of the electron's
wave-functions on a certain lattice site, its distribution captures the 
fluctuations of the wave-functions through the system.

The LD approach has been originally developed 
for a description of Anderson localization~\cite{abou73},
and furthermore been applied to systems with topological
disorder.~\cite{logan84,logan87}
In this article we like to demonstrate that it can also describe
systems with a bimodal disorder distribution, as the binary alloy model. 
It has been noted several times 
(see e.g. Refs.~\onlinecite{kirk72,taka74,souk92}), that 
the physics of this model is only partially covered in a mean field
description as provided by the coherent potential approximation
(CPA).
Instead one has to account for the specific configuration of atoms in the
vicinity of a lattice site.~\cite{dean58,tsukada69}
To give a better description, a variety of
(cluster) extensions to CPA has been devised, 
which explicitly treat correlations on finite clusters (see
e.g. Ref.~\onlinecite{taka74} or Ref.~\onlinecite{elliott74}).
We will show that the LD approach can serve as a conclusive extension of
CPA---indeed it contains
the CPA as a limit, see App.~\ref{appendix:LD}---which implicitly
contains these correlations. 

Besides giving very precise results for non-interacting systems,
an appealing feature of the LD approach
is its possible application to interacting disordered systems.  
Recently some progress in this direction has been made using the LD
approach in combination with dynamical mean field theory.~\cite{dobro98}
We could e.g. show that polaron formation in an
electron-phonon-coupled system is enhanced in the presence
of impurities, leading to polaron like defect states.~\cite{baf04}
The Mott transition in a binary alloy is another example,
where the LD approach might help to substantiate present
results.~\cite{byczuk04}

The outline of this article is as follows:
We will shortly introduce the binary alloy model and the associated
distributions, and then apply the LD approach.
As a limiting case we will consider the (quantum) percolation model,
and finally conclude. The two appendices contain the derivation of the
LD approach and its application to the Anderson localization problem.

\section{Model and distributions}\label{sec:modeldis}

The simplest model for an electron moving in a crystal with
substitutional disorder is given by the Hamiltonian
\begin{equation}\label{eq:hamilton}
H=\sum\limits_i \epsilon_i c^\dagger_i c^{\phantom{\dagger}}_i
- t
\sum\limits_{\langle i,j \rangle }
c^\dagger_i c^{\phantom{\dagger}}_j \; ,
\end{equation}
where $t$ denotes the tight-binding hopping integral between nearest
neighbour sites on a given lattice,
and the $\epsilon_i$'s are random on-site potentials.

The model is further specified through the probability
distribution of the random variables $\epsilon_i$.  
We only consider models where the 
$\epsilon_i$ are identically
independently distributed random variables
with a fixed distribution $p(\epsilon_i)$.
The Anderson model of localization, the binary alloy
model and the quantum percolation model are examples.

For given values of the $\epsilon_i$,
i.e. for a specific disorder configuration (`one specimen'),
the local density of states (LDOS)
\begin{equation}\label{def:LDOS}
\rho_i(\omega)=- \mathrm{Im}\; G_{ii}(\omega)/\pi
\end{equation}
 (expressed through the retarded
Green function $G_{ii}(\omega)$)
has a definite value on each lattice site $i$.
For the ordered system, translational symmetry implies that
$\rho_i(\omega)$ has the same value for every $i$.
In a disordered system, $\rho_i(\omega)$ varies with $i$,
so it takes on different values with a certain probability.
The corresponding probability distribution $p(\rho_i,\omega)$
captures the fluctuation of the LDOS through the system.
It is reasonable to assume that for an infinite system $p(\rho_i,\omega)$
does not depend on the explicit values of the $\epsilon_i$, but only
on the probability distribution $p(\epsilon_i)$.
We thus assume that the distribution
$p(\rho_i,\omega)$ is, in a slight `abuse of language',
self-averaging: It does not depend on the specific disorder configuration
looked at, but takes on a definite value in the thermodynamic limit. 
Moreover  $p(\rho_i,\omega)$ has a second meaning.
If we look at a fixed lattice site $i$
but consider all possible values of the $\epsilon_i$,
the LDOS is a random variable in its own right, whose probability
distribution is equal to $p(\rho_i,\omega)$ as defined above.  
So $p(\rho_i,\omega)$ gives 
(i) the probability that the LDOS has a certain value on some lattice
site, when all $\epsilon_i$ are fixed, and (ii) the probability
that the LDOS has a certain value on a fixed lattice site $i$, when
the $\epsilon_i$ vary.
Note that $p(\rho_i,\omega)$ does not depend on $i$, in contrast to $\rho_i(\omega)$.

One basic physical quantity that is calculated from
$p(\rho_i,\omega)$ is the 
\emph{arithmetically averaged density of states} (DOS)
\begin{equation}
 \rho(\omega) =  \int\limits_0^\infty
 \rho_i \; p(\rho_i,\omega) \; d\rho_i
\end{equation}
which counts the number of states at energy~$\omega$.
For the ordered system, when $\rho_i(\omega)$ does not depend on $i$,
$p(\rho_i,\omega)=\delta(\rho_i-\rho(\omega))$
is a $\delta$-peak at the DOS. 
In the presence of disorder 
$p(\rho_i,\omega)$ attains a certain width
and provides additional information on the
character of the electronic states in the system.
A narrow distribution corresponds to more or less homogeneous states,
when electron scattering is weak,
while a broad distribution reflects strong scattering leading to 
very inhomogeneous states.
The essential information on the character of states at energy $\omega$ is thus
provided by the distribution $p(\rho_i,\omega)$, which gives
a more detailed description of the disordered system
than the DOS $\rho(\omega)$ alone.

In particular we can decide whether states at $\omega$ are localized
if we employ the precise definition of $p(\rho_i,\omega)$.
The retarded Green function $G_{ii}(\omega)$, hence the LDOS and its
distribution, is usually calculated for a complex energy
$\omega+\mathrm{i}\eta$ with a small positive imaginary part $\eta$,
followed by analytical continuation to the real axis, i.e. $\eta\to0$.   
A finite $\eta$ gives a Lorentzian broadening with respect to
$\omega$, i.e. a finite energy resolution.
For $\eta\to0$, the resolution increases until peaks in the LDOS  
(corresponding to poles of $G_{ii}(\omega)$) and bands 
(corresponding to branch cuts of $G_{ii}(\omega)$) can be separated.
The behaviour of the distribution in the limit $\eta\to0$ is thus different depending
on the spectral properties of the system.
For a spectrum consisting only of discrete peaks---this corresponds
to localized states---the distribution becomes singular for $\eta\to0$.
Contrary to this we get a regular distribution if the spectrum is
continuous as for extended band (`Bloch') states.
We use this property in the study of Anderson localization (see
App.~\ref{appendix:anderson}), and will use it for the fragmented
spectra of the quantum percolation model.

In this article we obtain $p(\rho_i,\omega)$ directly through the
LD approach (for details see App. \ref{appendix:LD}).
It is natural to compare the results for $\rho(\omega)$
of the LD approach to the corresponding CPA results. 
While the lattice enters the CPA calculations only through the lattice
DOS for the ordered system the LD approach is explicitly constructed for a
Bethe lattice~\footnote{
The (half-infinite) Bethe lattice is a loop-free tree 
with a semicircular DOS 
$\rho(\omega)=(4/\pi W^2)\sqrt{W^2-4\omega^2}$.~\cite{economou83}
If not stated otherwise, we use a Bethe lattice with 
$K=2$ nearest neighbours to any lattice site.
The hopping matrix element $t$ is chosen to give a full
bandwidth $W=4 t \sqrt{K}=1$, so energies will be measured in units of $W$. 
}.
Owing to the absence of loops the Bethe lattice is a kind of 
mean-field approximation to (hyper-) cubic lattices.
Its particular geometry does not affect the
qualitative behaviour away from the localization transition, which
indeed is similar to a cubic lattice in three dimensions, see e.g. our
discussion of the quantum percolation model (Sec.~\ref{sec:perc}) or
the phase diagram obtained for the Anderson model
(App.~\ref{appendix:anderson}). Conversely the existence of different
phases and transitions in between is correctly described by the LD
approach for dimensions~$\ge 3$, similar to the success of dynamical
mean field theory for interacting systems in high dimensions.
Of course at the phase transition, when the correlation length
diverges, results for Bethe and cubic lattices differ 
(see Ref.~\onlinecite{mirlin94} and App.~\ref{appendix:anderson}). 
Weak localization in two dimensions, on the other hand,
is related to interference on closed loops
and thus not seen on a Bethe lattice.

\section{Binary Alloy Model}\label{sec:bam}

The binary alloy model has a bimodal disorder distribution
\begin{equation}
p(\epsilon_i)=c_A \delta(\epsilon_i-E_A)+(1-c_A)
\delta(\epsilon_i-E_B) \quad,
\end{equation}
corresponding to a crystal randomly composed of A-atoms (B-atoms) at energy
$E_A$ ($E_B$).
We set $E_A=-\Delta/2$, $E_B = \Delta/2$, i.e. $\Delta=E_B-E_A$.
Different aspects of the involved physics of the binary alloy model
have been discussed previously (see
e.g. Refs.~\onlinecite{kirk72,taka74,souk92,dean58,tsukada69}).  
The local electron motion
strongly depends on the configuration of atoms at (and in the vicinity of) a
certain lattice site. 
This `cluster effect', showing up as peaks in the DOS, is not obtained
by the CPA.
We will show that the LD approach can give a (more) complete picture with
fair ease. Especially the DOS becomes accessible and will
turn out to be important for a thorough understanding.

\subsection{Low A-concentration -- small separation energy}\label{subsec:bam1}
We first study the situation of low concentration
$c_A=0.1$ of A-atoms (which is below the classical percolation
threshold, cf. Sec. \ref{sec:perc}).
The A-atoms can be considered as the minority species in a bulk system
of B-atoms (doping a semiconductor is an example).

\begin{figure}[htb]
\begin{center}
\includegraphics[width=0.85\linewidth]{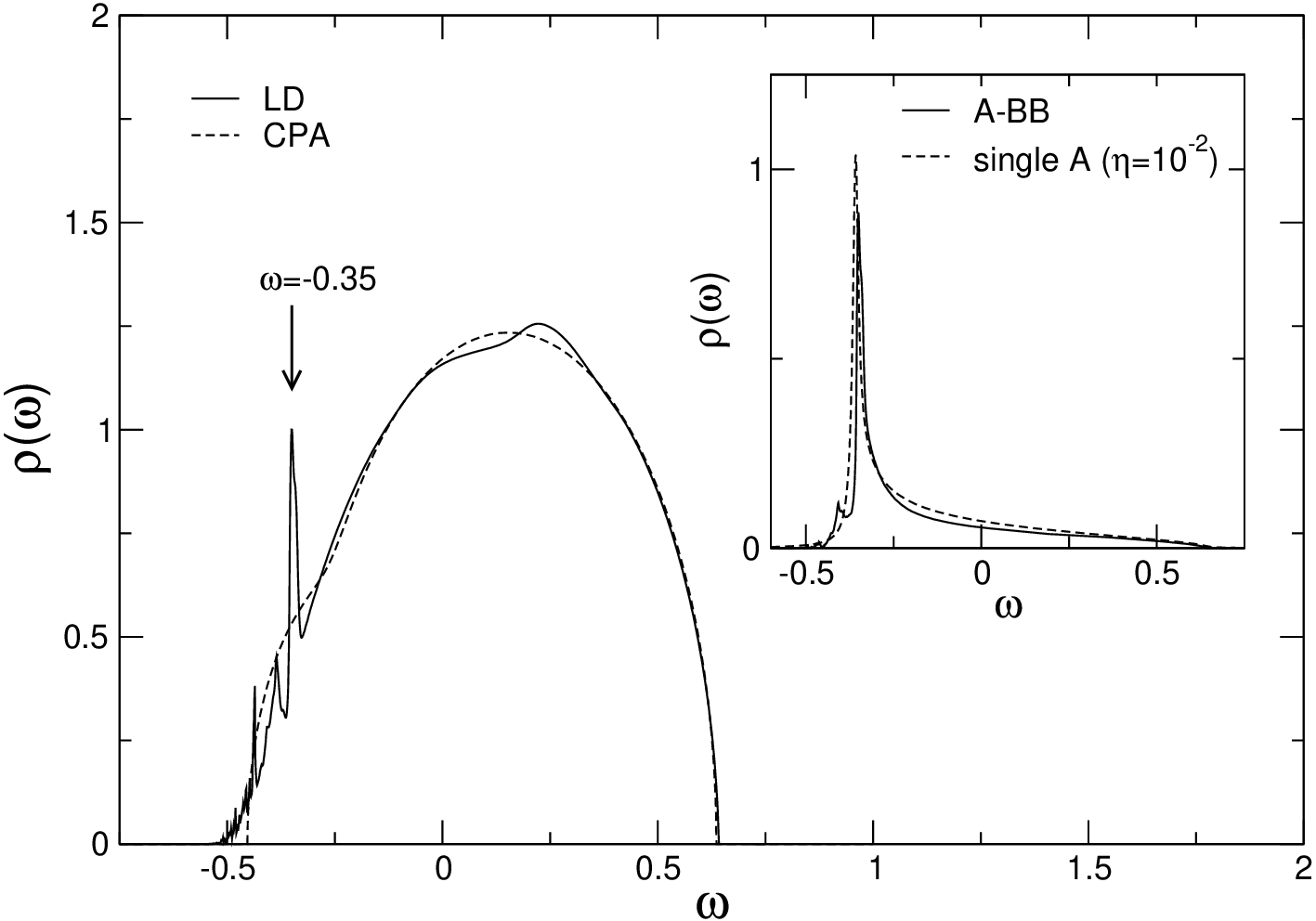}
%\hfill
\includegraphics[width=0.85\linewidth]{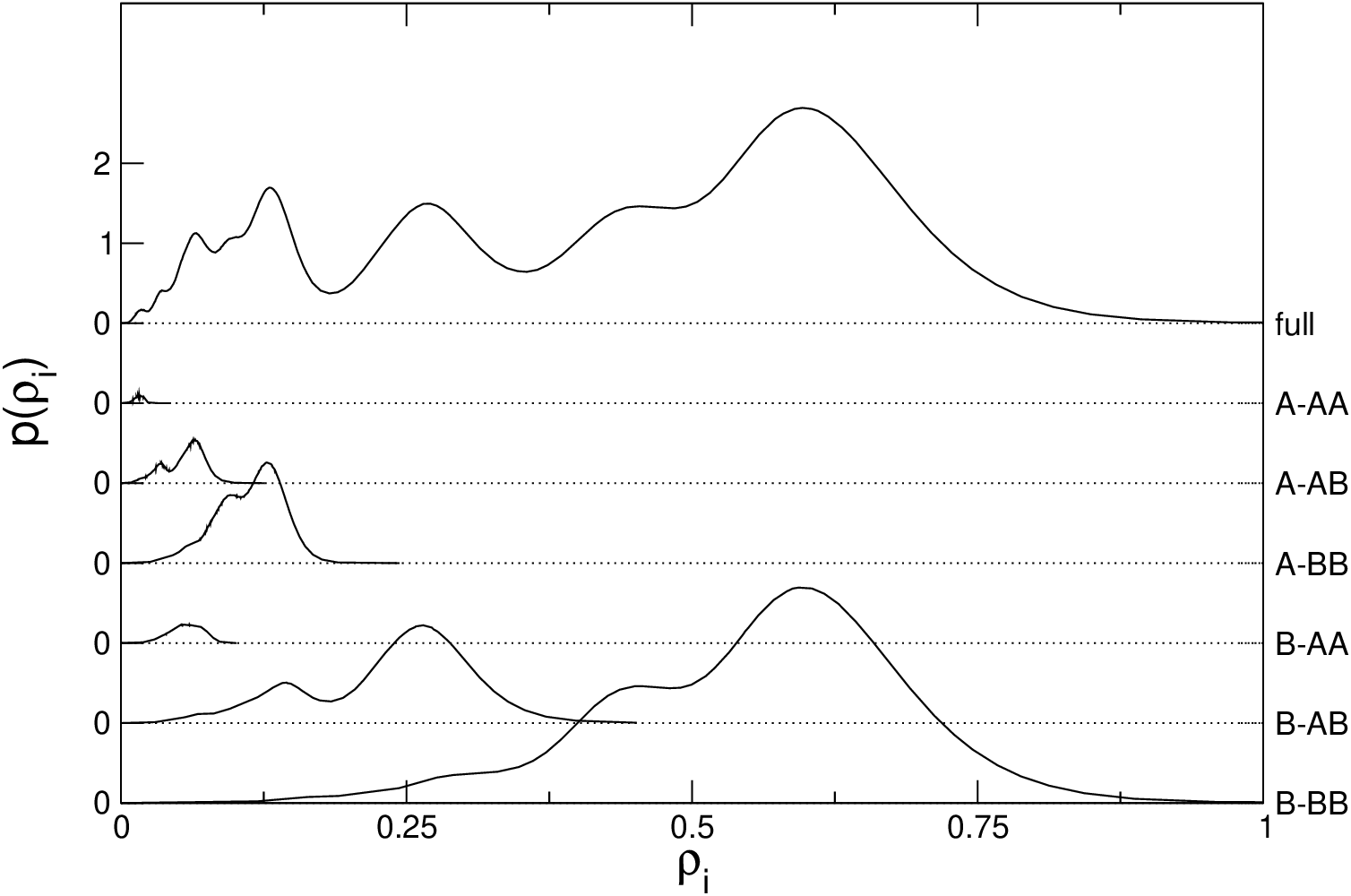}
\end{center}
\caption{
Binary alloy with $\Delta=0.3$, $c_A=0.1$.
The upper panel shows the  DOS, obtained within the LD approach and CPA.
The inset displays the conditional  DOS for the cluster configuration
$A-BB$, the dashed curve has been calculated using Eq.~(\ref{singleA}) with
broadening $\eta=10^{-2}$ (see text).
The lower panel shows the distribution of the LDOS at $\omega=0.6$
(curve on top). 
The six other curves show the distribution of the conditional LDOS for
a specified cluster of three sites.}
\label{fig:badosprob}
\end{figure}

For small separation energy $\Delta=0.3$
the energy levels of the minority A-atoms 
lie inside the B-band.
The DOS mainly consists of the B-band
centered at $E_B=+0.15$, but shows some additional structures at the lower
band edge,
which are absent in the CPA DOS
(cf. Fig.~\ref{fig:badosprob}).
They originate from the strong fluctuations in the local
environment of an atom,
which can be clearly seen in the distribution of the LDOS (cf. Fig.~\ref{fig:badosprob}).
Note the various peaks of this distribution (in contrast to the
Anderson model, see App.~\ref{appendix:anderson}), and its large width
displaying the pronounced fluctuations of the electron's wave-functions.

To understand the consequences of different atoms
situated at a lattice site
we can look at the `conditional' LDOS
$\rho^{A/B}_i(\omega)$,
subject to the constraint that an $A$-
respectively $B$-atom is located at site $i$.
More generally we can specify a certain configuration of atoms on
a cluster of sites centered at $i$.
By considering larger and larger clusters 
every peak in the distribution of the conditional LDOS (and in the DOS) can be attributed to a specific configuration
(cf. Fig.~\ref{fig:badosprob}). 
As the simplest example the pronounced peak at 
$\omega \approx -0.35$ results from a single A-atom surrounded by B-atoms
(cf. inset Fig.~\ref{fig:badosprob}).
Its approximate position and form
follow from the simple formula 
\begin{equation}\label{singleA}
  G_A(\omega)=\left[\frac{\omega+E_B}{2}-E_A+
\sqrt{\left(\frac{\omega-E_B}{2}\right)^2-\frac{W^2}{16}\;}
\; \right]^{-1} 
\end{equation}
for the local Green function of one A-impurity embedded in a
B-lattice. 
The corresponding DOS, using a Lorentzian broadening
$\eta=0.02$ in the energy argument $\omega+\mathrm{i}\eta$ of the
Green function to mimic the effect of tunnelling between different
A-atoms, fits the conditional DOS very well (dashed curve in
inset of Fig.~\ref{fig:badosprob}). 
The complementary situation of a B-atom surrounded by A-atoms contributes to
the `hump' in the B-band (cf. Fig.~\ref{fig:badosprob}),
which arises from B-atoms neighbouring to A-atoms.

It should be noted that Eq.~(\ref{singleA})
gives a $\delta$-peak outside the B-band,
corresponding to the impurity state at the A-atom,
only above a critical value of $\Delta$.
This is another evidence that the Bethe lattice should be understood as
an approximation to lattices in dimensions $d \ge 3$.

\subsection{Low A-concentration -- split band case}\label{subsec:bam2}

\begin{figure}[htb]
\begin{center}
\includegraphics[width=0.85\linewidth]{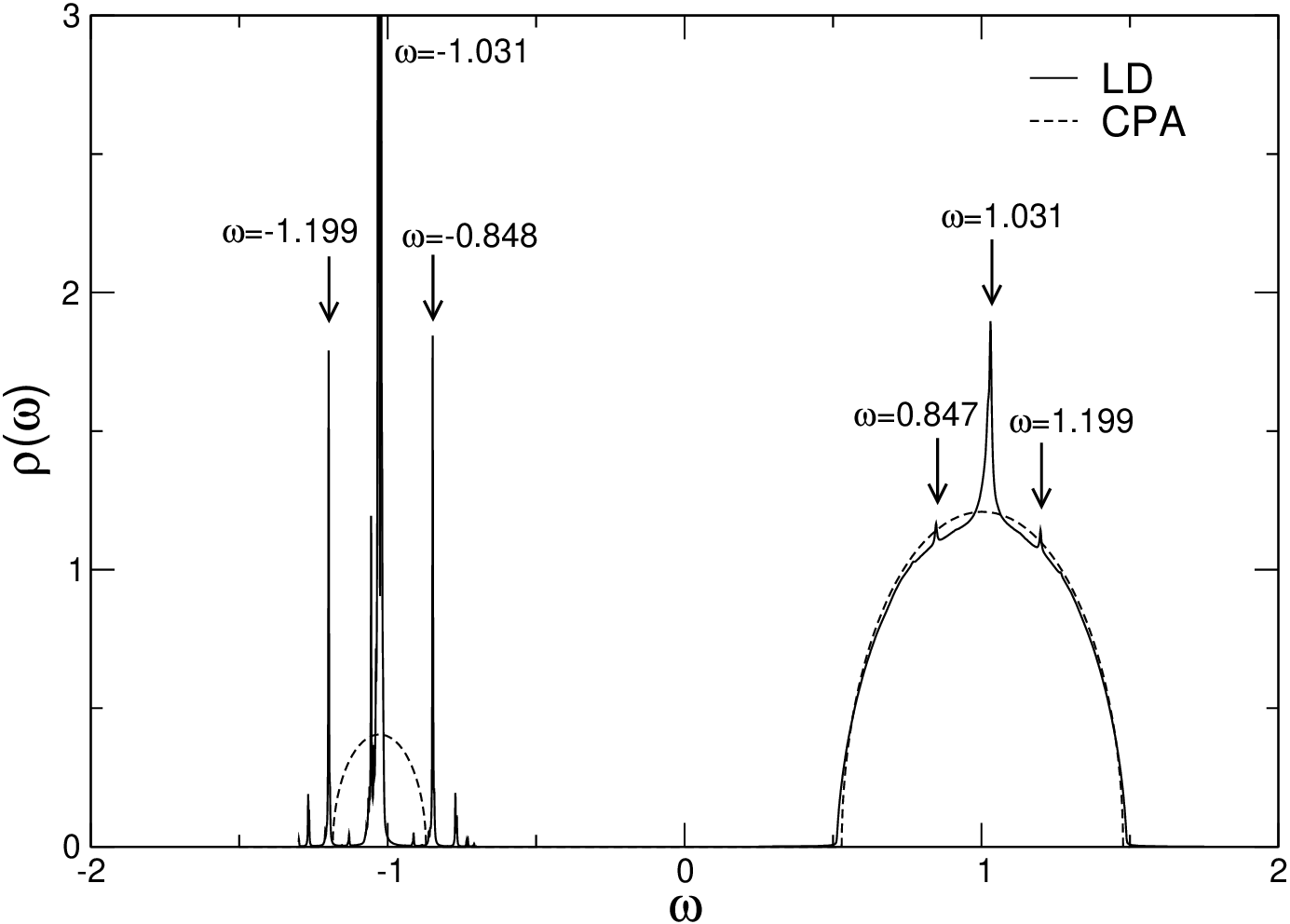}
\end{center}
\caption{
  DOS for the binary alloy with $\Delta=2.0$, $c_A=0.1$.
  To broaden the peaks of the `A-states'
  we add an artificial imaginary part $\eta=10^{-3}$ to the energy
  argument $\omega$ of the Green function $G_{ii}(\omega)$ in
  Eq.~(\ref{def:LDOS}). 
  Arrows mark the position of three peaks each in the A- and B-band.}
\label{fig:basplit}
\end{figure}

Increasing the separation energy to $\Delta=2.0$
(still with $c_A=0.1$) 
leads to the split band case (see Fig.~\ref{fig:basplit}).
The majority B-atoms still form a band whose DOS shows some additional spikes.

The concentration of the minority A-atoms is below the classical percolation
threshold hence only finite A-clusters exist (cf. Sec.~\ref{sec:perc}).
Due to the large energy separation the `A-states' on these clusters
are strongly damped through scattering on B-atoms.
Accordingly the `A-states' do not form a band of (extended)
states but a fragmented set of peaks with varying
height and width, reminiscent of the percolation model (cf. Sec. \ref{sec:perc}).
Again these peaks emerge from A-clusters embedded in the B-lattice.
The central peak at $\omega=-1.031$ 
whose position can be again calculated with Eq.~(\ref{singleA})
corresponds to a single A-atom surrounded by B-atoms.
Similarly the two side peaks result from two adjacent A-atoms, and so forth.
The weight of these peaks decreases exponentially as $c_A^N$
for a $N$-atomic cluster.
The complementary configurations, reversing the role of A- and B-atoms,
yield spikes in the B-band.
Since the concentration of B-atoms is large these spikes do not form
peaks but merge with the B-band.

Owing to the different strength of scattering we expect that
  all `A-states' are localized, while the `B-band' remains extended.
In accordance with localization in the Anderson model
(App.~\ref{appendix:anderson}) we can extract the nature of states from
the behaviour of the distribution $p(\rho_i,\omega)$
for $\omega$ shifted by a small imaginary part $\eta$.
Indeed we find that, for $\eta\to0$,
the distribution is singular in the energy range of the
`A-states' but regular for the B-band,
which proves that `A-states' are localized and the `B-band' remains
extended.

\subsection{Phase diagram}\label{subsec:bam3}

\begin{figure}[htb]
\begin{center}
\includegraphics[width=0.85\linewidth]{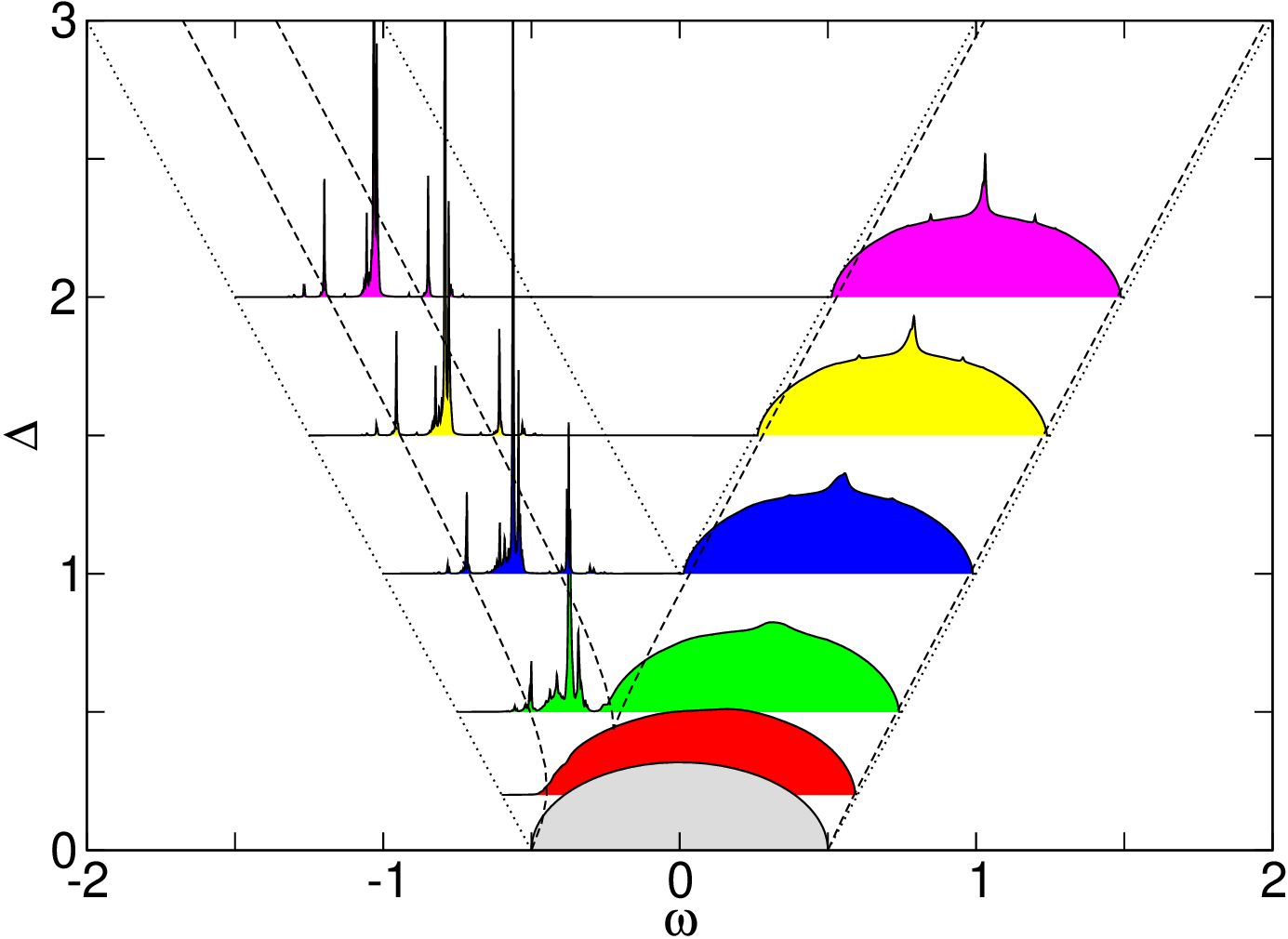}
\includegraphics[width=0.85\linewidth]{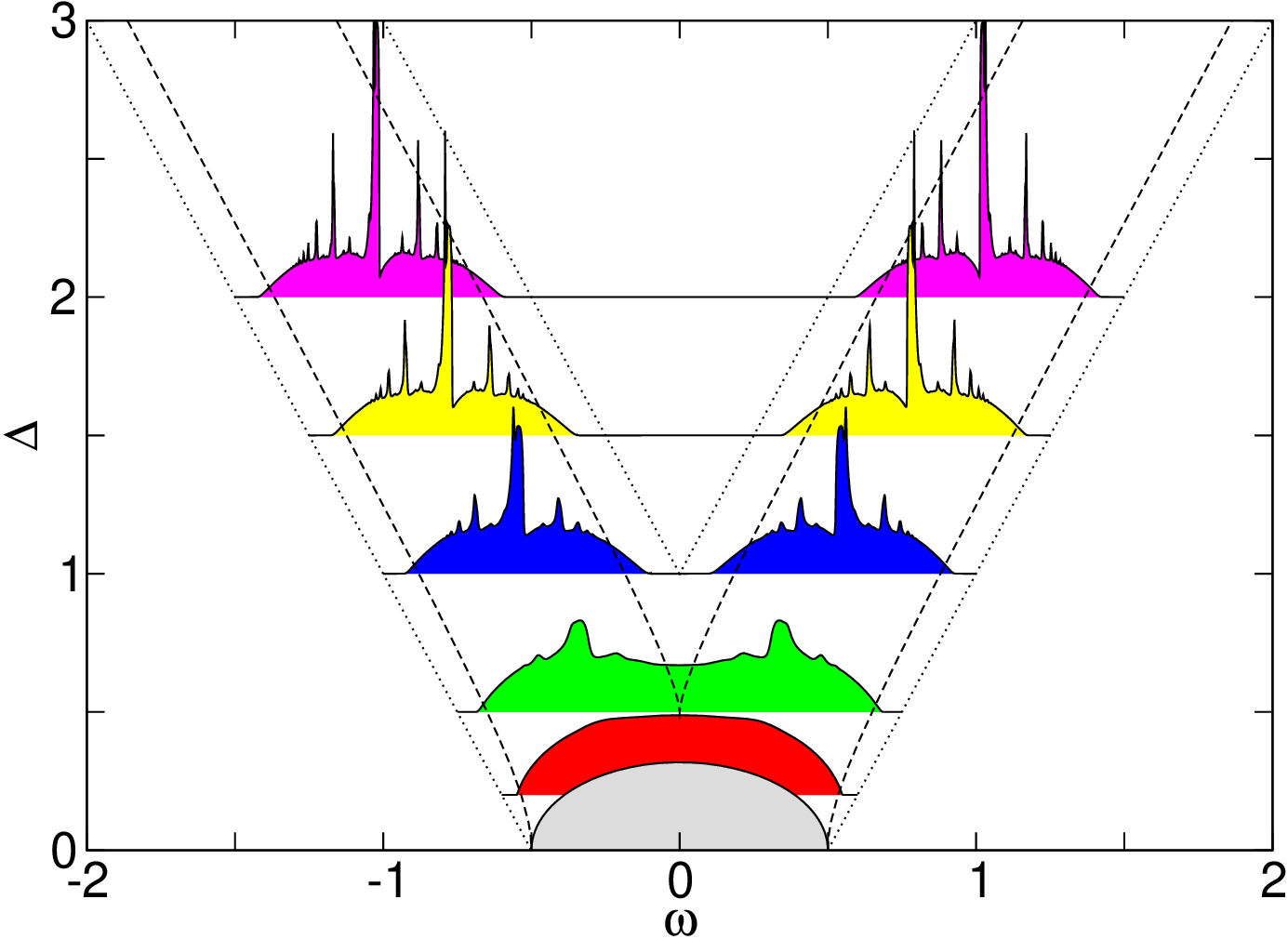}
\end{center}
\caption{\label{fig:baphase}
(Colour online) 
Phase diagram of the binary alloy,
for low concentration $c_A=0.1$ (upper panel),
and equal concentration $c_A=c_B=0.5$ (lower panel).
  The solid filled curves show the DOS for various $\Delta$.
  The dashed curves show the CPA band edges,
  and the dotted lines mark $\omega=\pm \Delta/2 \pm W/2$.}
\end{figure}

In Fig.~\ref{fig:baphase} we show the phase diagram of the binary
alloy for low concentration $c_A=0.1$. 
For $\Delta$ below a critical value $\Delta_c \approx 0.5 $ the spectrum
consists of a single band, which splits into two bands above $\Delta_c$.
The value of $\Delta_c$ is a little bit larger than obtained within CPA
($\Delta_c^\mathrm{CPA} \approx 0.45$), but significantly smaller than
$\Delta =1$, when two bands with bandwidth $W=1$ centered around
$\pm \Delta/2$ would no longer overlap.

\begin{figure}
\begin{center}
\includegraphics[width=0.85\linewidth]{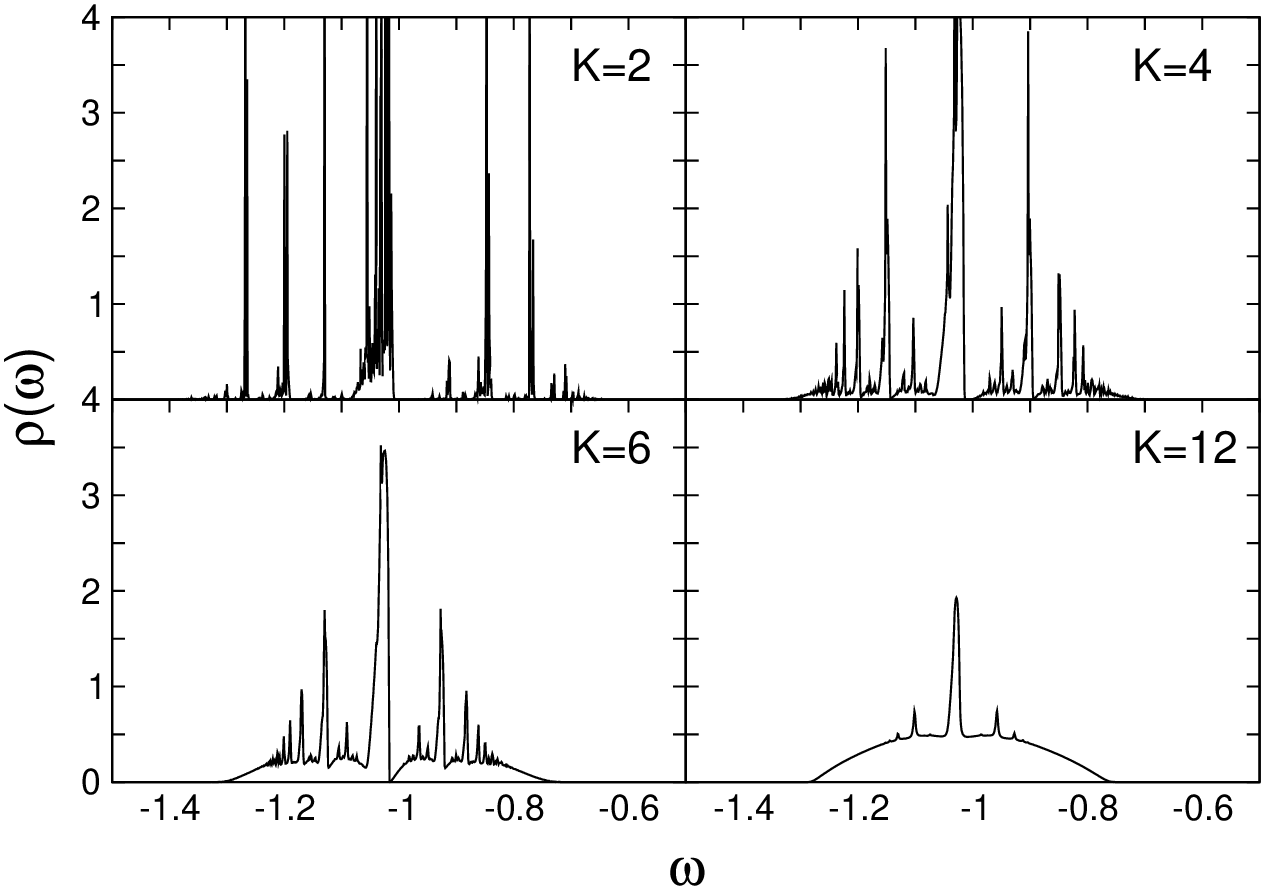}
\end{center}
\caption{\label{fig:IncK}
  DOS of the impurity band for $\Delta=2.0$ and $c_A=0.2$,
  with increasing number of nearest neighbours $K$ of the Bethe
  lattice.
  For $K=2,4$, the concentration $c_A$ is below the classical
  percolation threshold $p_c=1/K$, for $K=6,12$ above. 
  In contrast to Fig.~\ref{fig:basplit} we did not broaden the peaks
  with an additional $\eta$. }
\end{figure}

The phase diagram shows the rich physics of the binary alloy As
discussed in the previous two sections: 
(i) formation of many gaps, leading to a strongly fragmented minority `band',
(ii) the bands between the gaps show additional spikes,
(iii) for split bands and low concentrations,
minority states are localized impurity states,
while majority states are extended band states.
These three important effects go beyond the CPA,
which gives only the band splitting.
The CPA works reasonably well for the majority band,
where the local environment of a lattice site is less important,
but it cannot describe the fine structure of the minority band in even
a crude way.
These findings are in full agreement with Ref.~\onlinecite{souk92},
and we can additionally show the exact form of the DOS which
is one outcome of the LD approach.
So far we have dealt with a low concentration $c_A$ of the 
A-species below the classical percolation threshold $p_c = 1/K$ for
the Bethe lattice.
Then, when only finite A-clusters exist, strong signatures and
fragmentation of the DOS could be observed.
Increasing $K$ with $c_A$ fixed,
these signatures are partially weakened as long as $c_A < p_c$
(see Fig.~\ref{fig:IncK}).
Above the percolation threshold ($c_A > p_c$), infinite A-clusters
exist, leading to a band which is no longer fragmented.
Nevertheless the DOS still has peaks.
For large $K$ (equivalent to $p_c \ll c_A$) the DOS approaches a
semicircular form, anticipating the CPA result in the limit $K=\infty$
(cf. the discussion in the appendices).

For equal concentration $c_A=c_B=0.5$ and $K=3$, both atom species are
in the percolating regime, which corresponds not to a
doped material but to a stoichiometric compound.
The phase diagram (Fig.~\ref{fig:baphase}) shows that
the signatures in the DOS become more pronounced with increasing
$\Delta$, when the binary alloy model approaches
the percolation model (see Sec.~\ref{sec:perc}).

\section{Combined Anderson and binary alloy model}\label{sec:cab} 
\begin{figure}[htb]
\begin{center}
\includegraphics[width=0.85\linewidth]{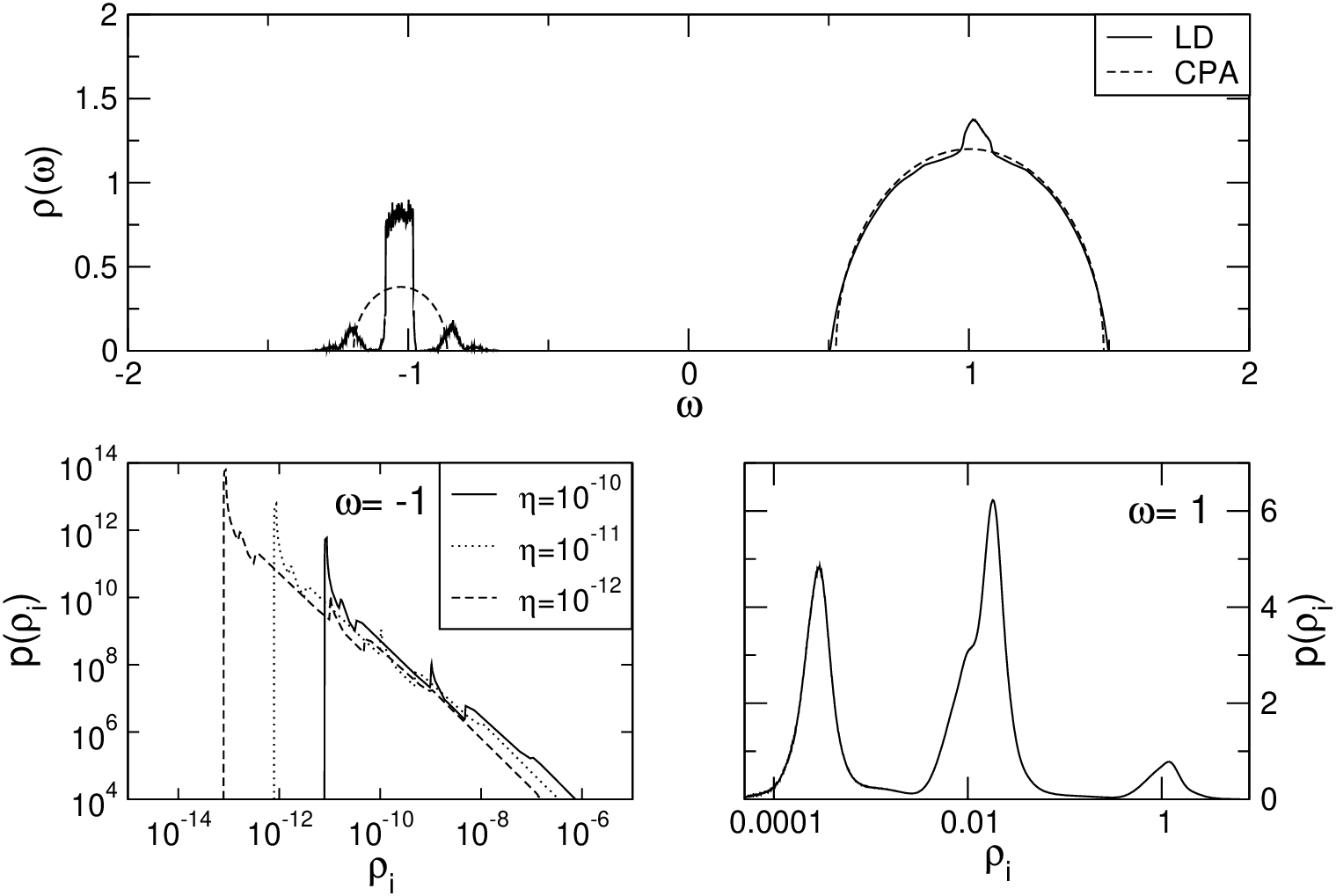}
\end{center}
\caption{\label{fig:bap}
The upper panel shows the DOS for the
binary alloy ($\Delta=2.0$, $c_A=0.1$) with additional on-site disorder
$\gamma=0.1$. (see Eq.~\ref{eq:bap}). 
The bottom row displays the dependence on $\eta$ of the LDOS
distribution.
Lower left panel: In the center of the A-band at $\omega=-1$
the distribution becomes singular for $\eta\to0$,
i.e. states are localized.
Lower right panel: For the B-band at $\omega=+1$ 
the distribution is independent of $\eta$,
and has a peaked structure known from the binary alloy model
(cf. Fig.~\ref{fig:badosprob}). }

\end{figure}

The binary alloy model shows a distinct tendency towards
peak and gap formation.
This behaviour is a generic feature of systems with a bimodal disorder
distribution.
For example consider the binary alloy
model with additional on-site box disorder $\gamma$, i.e.
\begin{equation}\label{eq:bap}
p(\epsilon_i) = \frac{c_A}{\gamma} \Theta(\gamma/2-|\epsilon_i-E_A|) 
+ \frac{1-c_A}{\gamma} \Theta(\gamma/2-|\epsilon_i-E_B|) \; .
\end{equation}
The CPA suggests that the system is the combination of two rescaled
Anderson models.
This is consistent with its phase diagram concerning the
position of mobility edges.~\cite{plyu03}
The LD approach additionally shows that
structures similar to the `pure' binary alloy appear
(cf. Fig.~\ref{fig:bap}). 
Evidently the system is not adequately described in terms
of rescaled Anderson models.
For the parameters considered $\gamma$ is on the same energy
scale as the hopping matrix element $t$ 
(here $t=W/\sqrt{32}$, i.e. $\gamma \approx 0.56 t$).
It is thus reasonable to assume that in a real alloy,
e.g. a doped semiconductor,
the peaked structure of the binary alloy DOS can be found.
Furthermore, if we extract localization properties of the system by the
$\eta\to0$-limit (cf. App.~\ref{appendix:anderson}), we find the A-band
to be entirely localized (see Fig.~\ref{fig:bap}).
Nevertheless $\gamma$ is much smaller than the critical disorder 
($\gamma_c \simeq 3 \times \mathrm{bandwidth}$) for Anderson
localization of the A-subband (in agreement with Ref.~\onlinecite{plyu03}). 
Therefore localization of the A-band is increased due to the strong
scattering occurring for the minority band in the binary alloy.
This suggests that impurity band states in a doped semiconductor will almost
always be localized. 

\section{Quantum Percolation Model}\label{sec:perc}

In the limit $E_B \to \infty$, now keeping $E_A=0$ fixed, the
binary alloy model reduces to the percolation model
where A-sites are embedded in an impenetrable host medium~\footnote{On
  the Bethe lattice site and bond percolation are equivalent.}.
Certainly, in a real alloy with $\Delta < \infty$,
the electron has the chance to tunnel through the host barrier.
Nevertheless, the percolation model 
shows very general features 
which have already shown up in the split band case of the binary alloy
model (see above). 

Let us first study the DOS.
If the concentration $c_A$ of A-atoms is below the classical
percolation threshold $p_c$ only finite clusters exist.
The corresponding spectrum is a pure point spectrum which densely fills
the energy interval $[-W/2,W/2]$ of the tight-binding band
(`Dirac comb').
An $N$-site cluster with occurrence probability
$c_A^N(1-c_A)^{N+1}$ contributes peaks at energies spread 
over the full possible range.
Consequently the weight of a peak varies
in contrast to the Anderson model non-monotonically with $\omega$.
Different statistical properties of the finite clusters can be
straightforwardly calculated on the Bethe lattice.
For instance the weight 
\begin{eqnarray}
  w_\mathrm{fin} &=& \sum\limits_{N=1}^\infty c_A^{N-1}(1-c_A)^{N+1}
  C_N   \\
  &=& 
\begin{cases}
 1, \quad &    c_A \le 0.5 \\
 (1-c_A)^2/c_A^2, &   c_A > 0.5
\end{cases}
\end{eqnarray}
of all finite clusters~\footnote{Again on a Bethe lattice with
  coordination number $K=2$, cf. first footnote.}
directly follows  
by help of the generating function for the Catalan numbers $C_N$, which give
the number of binary trees with $N$ sites.
For concentrations above $p_c \; (=0.5)$, when $w_\mathrm{fin}<1$,
an infinite percolating cluster exists.
This cluster can support extended states
which contribute to an continuous spectrum.
Since the `effective' dimension of the percolating cluster 
is smaller than $K$ (especially close to $p_c$),
the bandwidth of the resulting band is smaller than $W$.

\begin{figure}[htb]
\begin{center}
\includegraphics[width=0.85\linewidth]{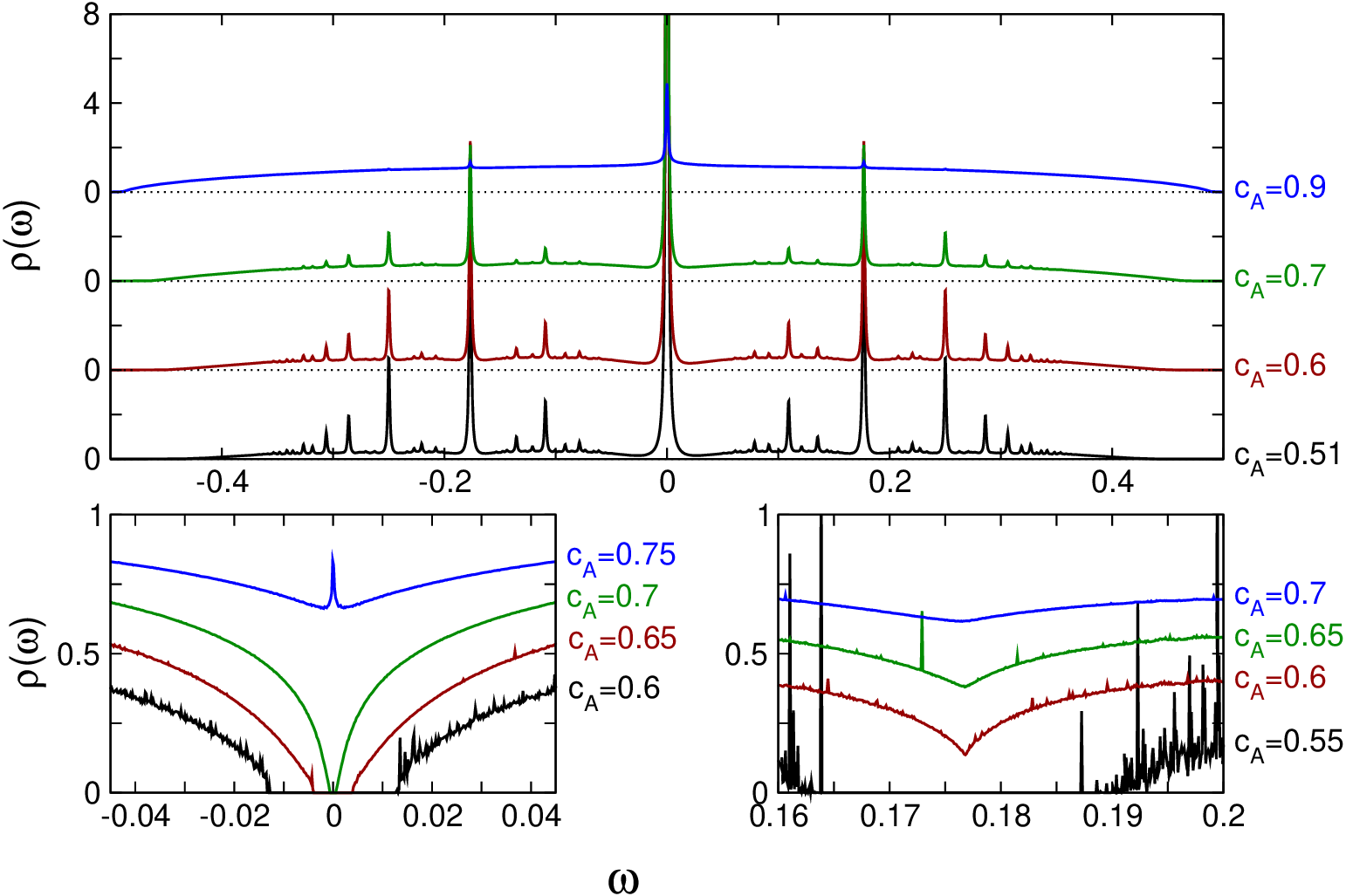}
\end{center}
\caption{\label{fig:percdos}
(Colour online)
  DOS $\rho(\omega)$ for the percolation
  model,
  for various concentrations $c_A$.
  Upper panel: Full spectrum with Lorentzian broadening $\eta=10^{-3}$.
Lower left panel: Central region around $\omega=0$ without Lorentzian
broadening.
Lower right panel: Region around the second peak at $\omega = t =
1/\sqrt{32} \approx 0.177$  without Lorentzian
broadening. }
\end{figure}

The DOS shows strong signatures (peaks and dips),
which have already occurred in the binary alloy model (see top picture in
Fig.~\ref{fig:percdos}). 
These signatures arise from both isolated finite clusters and 
finite clusters attached to the backbone of the infinite
percolating cluster.
With increasing concentration their weight reduces,
and the signatures are washed out.
At $\omega=0$ a pronounced $\delta$-peak surrounded by a
dip in the DOS exists.
With increasing concentration the peak reduces in weight and the dip
narrows, eventually both merge to a spike in the band.

The origin of this central peak and the dip
can be understood on the same level of reasoning as for
Eq.~(\ref{singleA}).
If a single atom is attached to the backbone of the percolating
cluster the Green function is modified by the additional hopping to this atom.
The DOS then shows a $\delta$-peak at 
 $\omega=0$, which is the energy of the state located at the atom,
and a dip around this peak which arises from damping of states on
the percolating cluster.
The same argumentation holds for any finite cluster instead of a
single atom.
Since larger clusters have lower probability of occurrence the central
peak is the most pronounced, and exists up to the largest
concentrations.

Nevertheless the percolating cluster is not made up of its backbone
plus one additional finite cluster, but many finite clusters attached
to it.
The Green function $G^c(\omega)$ of a (half-)infinite chain with one
additional atom attached to each site obeys the recursion relation 
\begin{equation}
G^c(\omega)=(\omega - t^2 /\omega - t^2 G^c(\omega)  )^{-1}
\end{equation}
which is solved by $G^c(\omega)=G^0(\omega-t^2/\omega)$,
where $G^0(\omega)$ is the Green function of the Bethe lattice.
Of course $G^c(\omega=0)=0$ as before,
but the diverging real part of $1/\omega$ produces not a dip but a \emph{gap}
in the DOS around $\omega=0$.
The validity of this simple argumentation
can be tested within the LD approach since its energy resolution
is not limited by finite size effects.
Without Lorentzian broadening, i.e. $\eta\to0$,
the $\delta$-peaks arising from finite cluster states
do not contribute to the DOS
(cf. Sec.~\ref{sec:modeldis}),
and only the continuous spectrum from extended band states survives.
This spectrum indeed shows a gap around $\omega=0$ for sufficiently
small concentrations (cf. Fig.~\ref{fig:percdos}, bottom left).
Lowering the concentration, gaps open at the energies corresponding
to any finite cluster eigenstate.
The first additional gaps open at $\omega=\pm t$
(Fig.~\ref{fig:percdos}, bottom right), corresponding to 
two site clusters, which have the second largest weight among all finite
clusters.
These gaps are filled by peaks from the dense spectrum of finite
cluster states, thus are absent in the DOS for any
finite energy resolution (cf. Ref.~\onlinecite{schubert05}).
The formation of gaps---which is a kind of `level repulsion'---is a
significant quantum feature with no counterpart in the classical
model. 

Besides the DOS the nature of states is
important.
For concentrations $c_A < p_c$, when only finite clusters exist,
all states are localized and no electron transport is possible.
For concentrations above $p_c$ a classical electron can traverse
the system along the percolating cluster. But a quantum mechanical
electron scatters off all irregularities and is possibly localized.
This raises the question of a quantum percolation threshold $p_q$
\emph{above}~$p_c$.

\begin{figure}[htb]
\begin{center}
\includegraphics[width=0.85\linewidth]{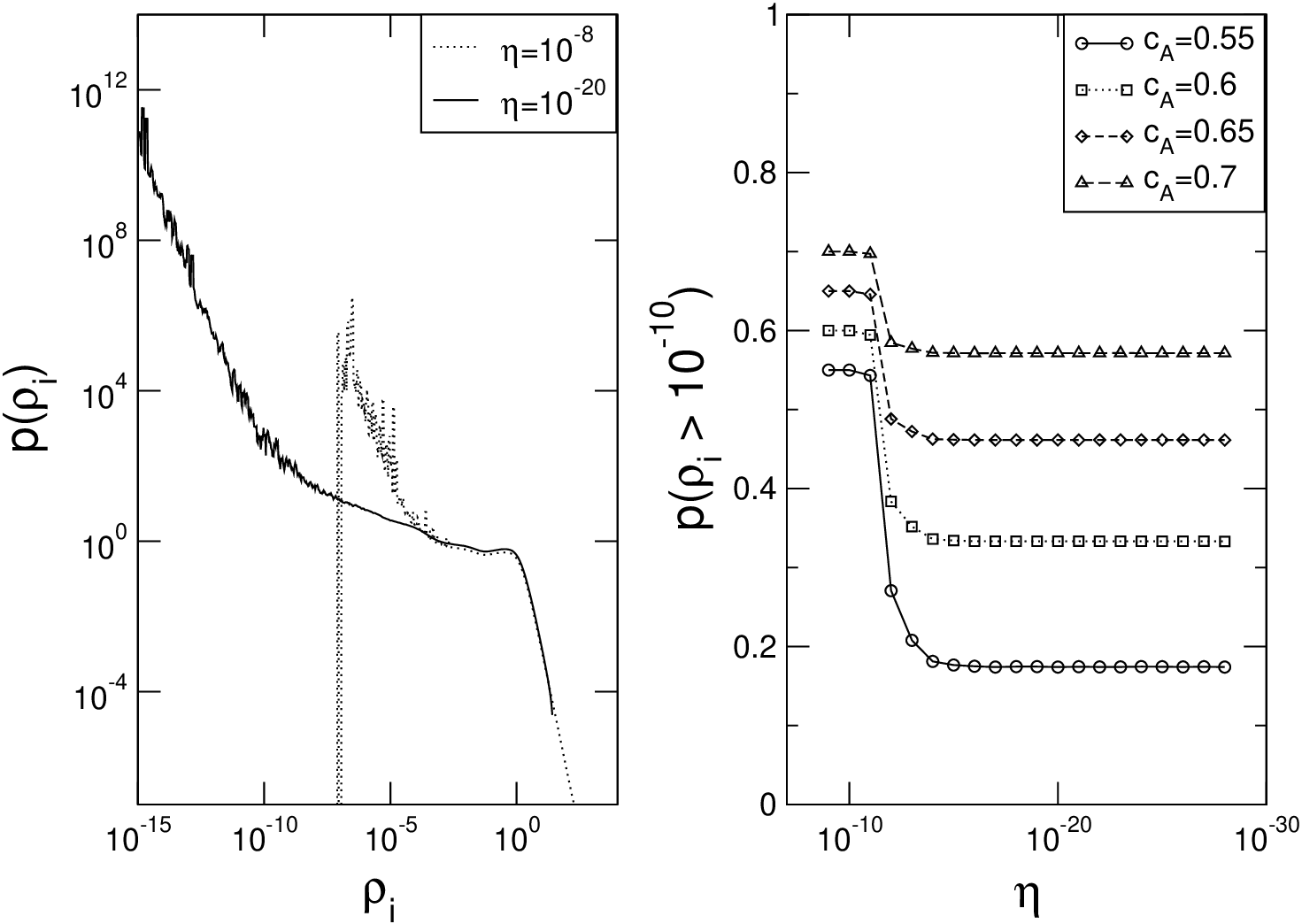}
\end{center}
\caption{\label{fig:perchisto}
LDOS distribution for the percolation model at $\omega=0.1$. 
  Left panel: For $c_A=0.7$ and two values of regularization
  $\eta=10^{-8}$ and $\eta=10^{-20}$. 
  For $\eta\to0$
 only extended states contribute to $p(\rho_i,\omega)$ at finite $\rho_i$.
  Right panel: 
  Integrated probability 
  \mbox{$p(\rho_i > \xi,\omega)$}
 in dependence on $\eta$,
  for various $c_A$ and one particular value $\xi=10^{-10}$. 
  For $c_A \lesssim 0.54$ no states exist at $\omega=0.1$,
  and $p(\rho_i,\omega)=0$. }
\end{figure}

Close to $p_c$ states on the percolating cluster have only
exponentially small weight. It is thus difficult to extract the
information about those states from the background of finite cluster
contributions.
However, we can gain some information from the distribution of the
LDOS, if $c_A$ and $\eta$ are varied.
Again, if we let $\eta\to0$, the energy resolution increases until
contributions from extended states on
the percolating cluster are separated from the discrete peaks of
localized states (see left panel of Fig.~\ref{fig:perchisto}).
For fixed energy $\omega$ and suitably large concentration $c_A$
the distribution $p(\rho_i,\omega)$ has finite weight at
finite $\rho_i$ in the limit $\eta\to0$ (see right
panel of Fig.~\ref{fig:perchisto}), which
indicates that extended states exist. 
Their weight can be estimated through the integrated probability
$p(\rho_i > \xi,\omega)=\int_\xi^\infty p(\rho_i,\omega) \;d\rho_i$,
with $\eta < \xi$.
The weight decreases with $c_A$,
and abruptly drops to zero at a certain concentration (for $c_A
\approx 0.54$ at $\omega=0.1$). 
Then no states exist at $\omega$, and $\rho_\mathrm{ave}(\omega)=0$.
This behaviour has to be attributed to the fragmentation of the spectrum
discussed before.
At a given energy $\omega$ a gap will open for concentrations $c_A$
sufficiently close to $p_c$. 
Before a gap opens at $\omega$,
states are extended.
After the gap has opened, no states at $\omega$ exist.
Between the gaps, which open at different
concentrations for different energies,
 extended states can survive even for very small
$c_A$, although all states are extremely damped and practically localized.
A definite localization transition from extended to localized states
does however not take place, and the only transition occurs when the
spectrum is fully fragmented, that is at the classical transition $c_A=p_c$. 
We conclude, on the basis of these arguments and our numerical
results, that, for the percolation problem on the Bethe lattice, a
quantum percolation threshold above the classical one does not exist. 
Note that scattering on the percolating cluster is of
a different type than for the Anderson model.
The finite clusters attached to the percolating backbone do not act
as coherent but incoherent scatterers.
So states on the backbone will not be localized for small $c_A$
(i.e. strong scattering), or even immediately localized, as in one
dimension.

\section{Conclusions}\label{sec:concl}

In this article we demonstrated how the 
LD approach can be used to study 
localization and percolative effects in alloys.
With very moderate computational demands
this scheme suffices to resolve the rich structures
in the DOS originating from comparably strong disorder
fluctuations.
Even for the extreme limit of the binary alloy model,
the percolation model, convincing results are easily obtained.
For instance the question whether gaps form in the percolation model
could be definitely answered,
and the possibility of a quantum percolation threshold above the
classical one could be almost definitely ruled out for the Bethe lattice.

We conclude that the LD approach is a convenient framework
for investigations of disorder and localization, and suggest
its application to interacting disordered systems.

\appendix

\section{The LD approach}\label{appendix:LD}
 
The LD approach has been constructed by Abou-Chacra, Anderson, and
Thouless~\cite{abou73} on a Bethe lattice.
There the local Green function 
$ G_{ii}(\omega) = \langle i | (\omega + \mathrm{i}\eta -H)^{-1} |i \rangle$
can be expressed through Green functions on the $K$ neighbouring lattice
sites $j=1,\dots,K$,
\begin{equation} \label{LDeq}
 G_{ii}(\omega) = \Bigl[\omega-\epsilon_i-
 t^2 \sum\limits_{j=1}^K G_{jj}(\omega) \Bigr]^{-1}
 \quad ,
\end{equation}
where each $G_{jj}(\omega)$ on the r.h.s. of this
equation is evaluated for the lattice with site $i$ removed.
Iterating this expansion an infinite hierarchy of equations is generated
(`renormalized perturbation expansion').~\cite{economou83}
Instead of solving this hierarchy for many particular realizations
of the $\epsilon_i$ and constructing the distribution of
$G_{ii}(\omega)$ afterwards
the LD approach manages to solve Eq.~(\ref{LDeq}) directly for the
distribution. 
This solution relies on two properties arising from the special
geometry of the Bethe lattice.
First, all Green functions in Eq.~(\ref{LDeq}) correspond to the same
geometric situation (one lattice site with $K$ neighbouring sites).
Hence their distribution is identical (although 
their concrete values differ).
Moreover, removing site $i$ from the lattice, the lattice sites
$j=1,\dots,K$ are unconnected, and the $G_{jj}(\omega)$
are independently distributed.
Owing to these two properties 
Eq.~(\ref{LDeq}) expresses one random variable through $K$
independently distributed random variables with the same distribution,
and can therefore be interpreted
as a self-consistency equation for the distribution
of $G_{ii}(\omega)$.~\cite{abou73,baf04}
Be aware that then the indices $i$, $j$ do not denote specific lattice
sites but certain realizations of the random variable
$G_{ii}(\omega)$.

For the ordered system ($\epsilon_i=0$) 
all Green functions in Eq.~(\ref{LDeq}) are identical,
and the Green function
$G^0(\omega)= (8/W^2)(\omega - \sqrt{\omega^2-W^2/4})$ of the Bethe
lattice~\cite{economou83}, with bandwidth $W=4 t \sqrt{K}$, is
obtained.
For a disordered system the solution of
Eq.~(\ref{LDeq}) is obtained through a Monte-Carlo procedure (Gibbs
sampling).
The distribution is represented through a sample of typically
$10^4$ up to $10^7$ elements, depending
on the respective case studied.
At each step of iteration a new sample is constructed
whose elements are calculated through Eq.~(\ref{LDeq})
with a randomly chosen $\epsilon_i$ and $K$ elements drawn from the
previous sample. A hundred up to some thousand
iterations are necessary to guarantee convergence.
The resulting computation time on a standard desktop PC ranges
from few minutes to some hours.

The LD approach comprises the coherent potential approximation (CPA) in
the limit of infinite coordination number $K=\infty$.
Taking into account the scaling $t \propto
1/\sqrt{K}$ of the hopping matrix element, the sum over $j$ in
Eq.~(\ref{LDeq}) can---assuming the central limit
theorem to be applicable---be replaced by the arithmetic average of
$G_{jj}$, and the CPA is recovered.

In course of its construction the LD approach works on an infinite lattice
(no boundaries, no finite size effects, no finite energy resolution).
In the numerical solution the size of the sample determines the
resolution which the distribution is sampled with. The resolution can
be easily enhanced by increasing the sample size (see below).

\section{Anderson model}\label{appendix:anderson}

We describe in this
appendix the application of the LD approach to the Anderson
localization problem. 
Although this is not the main objective of this article,
it will help to underline the generality of the LD approach.

The Anderson model---which is the prototype model showing a
localization transition---is given by Eq.~(\ref{eq:hamilton}) for
a uniform distribution of $\epsilon_i$ in the interval
$[-\gamma/2,\gamma/2]$,
\begin{equation}
 p(\epsilon_i) =
 \frac{1}{\gamma}\Theta\left(\frac{\gamma}{2}-|\epsilon_i|\right) 
\quad.
\end{equation}

\begin{figure}[htb] 
\begin{center}
\includegraphics[width=0.85\linewidth]{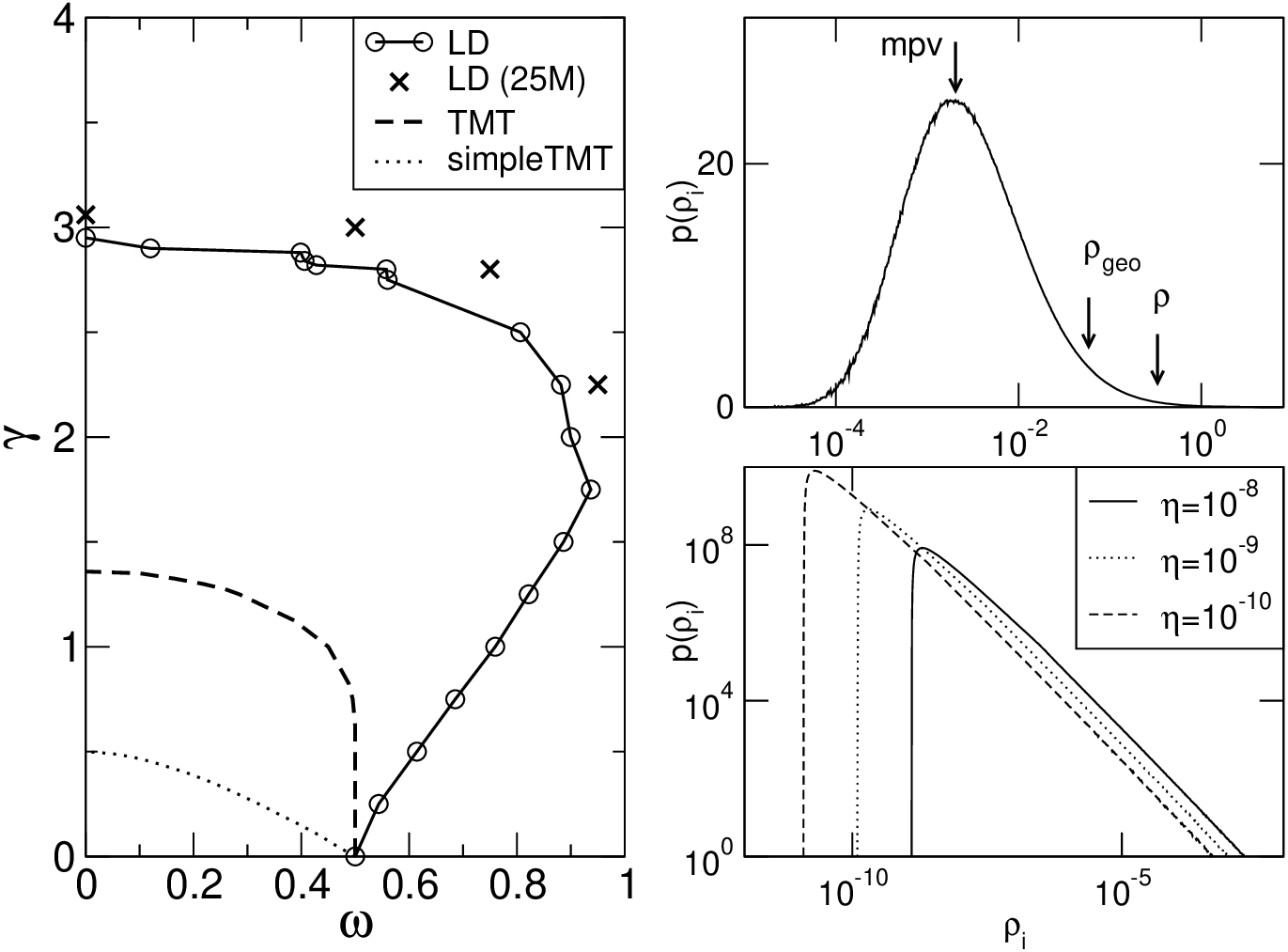}
\end{center}
\caption{Left panel: Phase diagram of the Anderson model.
  The solid (dashed, dotted) curve shows the mobility edge trajectory
  calculated within the LD approach (TMT, simplified TMT) by means of the
  $\eta\to0$ limit, using  a sample with  $5\times 10^4$ elements.
  The crosses indicate points in the $(\omega,\gamma)$-plane
  corresponding to the mobility edge position for an extremely large sample of
  $2.5\times10^7$ elements. 
  Upper right panel: probability distribution of LDOS for extended states,
at $\gamma=1.5$, $\omega=0.0$.
The distribution is -- for not too large values of $\eta$ -- 
independent of $\eta$, but the most probable value (`mpv') is some
  orders of magnitude smaller than the (geometric) DOS $\rho(\omega)$
  ($\rho_\mathrm{geo}(\omega)$).
Lower right panel: probability distribution $p(\rho_i,\omega)$ for
  localized states, 
at $\gamma=1.5$, $\omega=0.9$, and three values of~$\eta$.
}
\label{fig:Anderson}
\end{figure}

The characteristics of localization  
show up in $p(\rho_i,\omega)$
(cf. Fig.~(\ref{fig:Anderson})).
Impurity scattering causes the distribution to be strongly asymmetric
and broad even for extended states.
As one consequence the DOS $\rho(\omega)$ is on a different
scale than `typical' values of the distribution, e.g. the most
probable value. 
On approaching the localization transition the asymmetry further increases.
Much weight is transferred to large values of $\rho_i$, while the
`typical' values tends to zero.

A closer look at the distribution reveals 
that near the localization transition 
it shows a power law behaviour
over a wide range of $\rho_i$,
with an exponent $\simeq 1.45$, which is reasonably close
to analytical results obtained from field theoretical
considerations.~\cite{mirlin94}

Passing the localization transition the distribution becomes singular,
corresponding to a transition from continuous to discrete spectrum.
This characteristic change reflects itself most clearly in the
dependence of the distribution on the $\eta$-regularization,
when $G_{ii}(\omega)$ is calculated for a complex energy argument
$\omega+\mathrm{i}\eta$. 
While the distribution for extended states is stable when decreasing $\eta$, 
the relevant scale for localized states is entirely set through $\eta$,
and the distribution becomes singular for $\eta\to 0$ 
(see Fig.~\ref{fig:Anderson}). 
We can use this different behaviour as a localization criterion,
if we perform the $\eta\to0$ limit numerically.
This criterion does not depend on any a priori choice or approximation,
hence should be considered `numerically exact'.
As we have already mentioned (cf. App.~\ref{appendix:LD})
the finite size of the Monte-Carlo sample
sets the resolution for sampling $p(\rho_i,\omega)$.
If the sample is too small
(and thus the resolution too low) 
an almost singular distribution will be falsely detected as
singular while a larger sample correctly gives a regular distribution
corresponding to extended states.
Accordingly the mobility edge is shifted to larger
values of disorder on increasing the sample size (Fig.~\ref{fig:Anderson}).
However, the points on the mobility edge trajectory
readily stabilize if the sample is chosen large enough,
and a precise determination of the mobility edge is possible.

The LD approach phase diagram for the Anderson model which is then obtained
shows the characteristic features of the
localization problem in three dimensions.~\cite{kramer93,schubert04}
These can be most simply understood
to arise from the interplay of two competing effects:
While, for small disorder, tunnelling between shallow impurities produces
extended states outside the tight-binding band,
strong scattering on deep impurities takes place with
increasing disorder, starting to localize formerly extended states.
Therefore a reentrant behaviour of the mobility edge trajectory and
the existence of a critical disorder $\gamma_c$ 
($\gamma_c \approx 3.0$ for $K=2$ neighbour sites
of the Bethe lattice) for complete localization of all states is found. 

According to the mean-field type of approximation, the critical
disorder $\gamma_c$ for the Bethe lattice is larger than
for a cubic lattice.
For $K\to\infty$, $\gamma_c$ grows without bound,
i.e. localization is absent in $K=\infty$ (where
the LD approach reduces to the CPA).
For $K=1$ the Bethe lattice is a one-dimensional chain, where all
states are known to be localized for arbitrary disorder.
The sum in Eq.~(\ref{LDeq}) then contains only one $G_{jj}(\omega)$,
which implies that the LD sampling scheme cannot
converge to a stable distribution, since different elements of the
sample never become related to each other during the sampling.
This instability expresses the particular one-dimensional localization
behaviour.

For comparison to the LD approach we show in Fig.~\ref{fig:Anderson} 
the mobility edge trajectory obtained
within a recently proposed mean-field like approach to Anderson
localization, the so-called typical medium theory
(TMT).~\cite{dobro02,byczuk05}
This TMT modifies the CPA
by reformulating its self-consistency
condition in terms of the geometrically averaged
DOS
\begin{equation}
 \rho_\mathrm{geo}(\omega) = \exp
 \left(\int\limits_0^\infty (\ln \rho_i ) p(\rho_i,\omega) d\rho_i \right)
\quad.
\end{equation}
This average is known to be critical at the localization transition,
since it puts much weight at low values of $\rho_i$.
It nevertheless does not approximate the `typical' values,
see Fig.~\ref{fig:Anderson}. 

If we compare the lines of vanishing $\rho_\mathrm{geo}(\omega)$ from TMT
(i.e. the `TMT mobility edges') with
the LD approach phase diagram we see the consequences of this modification:
(i) the critical disorder predicted is significantly smaller,
and (ii) the reentrant behaviour of the mobility edge is entirely missed.

Remember that the CPA is obtained in the well defined limit of
infinite coordination number $K=\infty$, therefore is a controlled
approximation.
In the TMT construction some ambiguity enters in the choice of the
average used.
In fact TMT can be further simplified,
replacing $\rho_\mathrm{geo}(\omega)$ by
$\rho_\mathrm{simp}=\mathrm{min}\{\rho_i(\epsilon_i=\gamma/2),\rho_i(\epsilon_i=-\gamma/2)\}$.
This `averaging procedure' drastically overestimates the strength of
impurity scattering,
and the `simplified TMT' is surely far away from any reasonable
description of the underlying physics.
However, the phase diagram obtained is very similar to the TMT one.
Of course the critical disorder has to be even smaller than in TMT.

Apparently TMT captures strong impurity scattering  
which is partially neglected in the CPA.
However, Anderson localization is not merely a result
of strong scattering but quantum interference due to coherent scattering.
The similarity between the TMT and `simplified TMT' results indicates
that TMT might not adequately include these interference effects.


\begin{thebibliography}{10}

\bibitem{anderson58}
P.~W. Anderson,
\newblock Physical Review {\bf 109}, 1492 (1958).

\bibitem{abou73}
R.~Abou-Chacra, P.~W. Anderson, and D.~J. Thouless,
\newblock J.\ Phys.\ C {\bf 6}, 1734 (1973).

\bibitem{logan84}
D.~E. Logan and P.~G. Wolynes,
\newblock Phys.\ Rev.\ B {\bf 29}, 6560 (1984).

\bibitem{logan87}
D.~E. Logan and P.~G. Wolynes,
\newblock Phys.\ Rev.\ B {\bf 36}, 4135 (1987).

\bibitem{kirk72}
S.~Kirkpatrick and T.~P. Eggarter,
\newblock Phys.\ Rev.\ B {\bf 72}, 3598 (1972).

\bibitem{taka74}
I.~Takahashi and M.~Shimizu,
\newblock Prog.\ Theor.\ Phys. {\bf 51}, 1678 (1974).

\bibitem{souk92}
C.~M. Soukoulis, Q.~Li, and G.~S. Grest,
\newblock Phys.\ Rev.\ B {\bf 45}, 7724 (1992).

\bibitem{dean58}
P.~Dean,
\newblock Proc.\ Phys.\ Soc. {\bf 73}, 413 (1959).

\bibitem{tsukada69}
M.~Tsukada,
\newblock J.\ Phys.\ Soc.\ Japan {\bf 26}, 684 (1969).

\bibitem{elliott74}
R.~J. Elliott, J.~A. Krumhansl, and P.~L. Leath,
\newblock Rev.\ Mod.\ Phys. {\bf 46}, 465 (1974).

\bibitem{dobro98}
V.~Dobrosavljevi\'{c} and G.~Kotliar,
\newblock Phil.\ Trans.\ R.\ Soc.\ Lond.\ A {\bf 356}, 57 (1998).

\bibitem{baf04}
F.~X. Bronold, A.~Alvermann, and H.~Fehske,
\newblock Phil.\ Mag. {\bf 84}, 673 (2004).

\bibitem{byczuk04}
K.~Byczuk, W.~Hofstetter, and D.~Vollhardt,
\newblock Phys.\ Rev.\ B {\bf 69}, 045112 (2004).

\bibitem{mirlin94}
A.~D. Mirlin and Y.~V. Fyodorov,
\newblock Phys.\ Rev.\ Lett. {\bf 72}, 526 (1994).

\bibitem{plyu03}
I.~V. Plyushchay, R.~A. R{\"o}mer, and M.~Schreiber,
\newblock Phys.\ Rev.\ B {\bf 68}, 064201 (2003).

\bibitem{schubert05}
G.~Schubert, A.~Wei{\ss}e, and H.~Fehske,
\newblock Phys.\ Rev.\ B {\bf 71}, 045126 (2005).

\bibitem{economou83}
E.~N. Economou,
\newblock {\em Green's Functions in Quantum Physics},
\newblock Springer-Verlag, Berlin, 1983.

\bibitem{kramer93}
B.~Kramer and A.~MacKinnon,
\newblock Rep. Prog. Phys. {\bf 56}, 1469 (1993).

\bibitem{schubert04}
G.~Schubert, A.~Wei{\ss}e, G.~Wellein, and H.~Fehske,
\newblock in {\em High Performance Computing in Science and
  Engineering, Garching 2004},
\newblock edited by A.~Bode, F.~Durst (Springer-Verlag, Berlin,
Heidelberg, 2005), pp. 237--250.

\bibitem{dobro02}
V.~Dobrosavljevi\'{c}, A.~A. Pastor, and B.~K. Nikoli\'{c},
\newblock Europhys. Lett. {\bf 62}, 76 (2003).

\bibitem{byczuk05}
K.~Byczuk, W.~Hofstetter, and D.~Vollhardt,
\newblock Phys.\ Rev.\ Lett. {\bf 94}, 056404 (2005).

\end{thebibliography}
\end{document}